\documentclass[twocolumn,showpacs,aps,prd,superscriptaddress,preprintnumbers,amsmath,amssymb,floatfix]{revtex4}

\usepackage{graphicx}
\usepackage{dcolumn}
\usepackage{bm}

\usepackage{relsize}
\usepackage{xspace}
\def\babar{\mbox{\slshape B\kern-0.1em{\smaller A}\kern-0.1em
    B\kern-0.1em{\smaller A\kern-0.2em R}}}
\def\pep2{PEP-II}

\def\Km{K^-}
\def\Kp{K^+}
\def\piz{\pi^0}
\def\pip{\pi^+}

\def\Ds{D^+_s}

\def\DsFE{D_{sJ}(2458)^+}
\def\DsTT{D_{sJ}^*(2317)^+}
\def\DsTO{D_s^{*}(2112)^+}
\def\Kbar{\kern 0.2em\overline{\kern -0.2em K}{}\xspace}

\def\DmDp{\Delta m_{\piz}}
\def\DmDg{\Delta m_\gamma}

\newcommand{\gevc}{\ensuremath{{\mathrm{\,Ge\kern -0.1em V\!/}c}}\xspace}
\newcommand{\mevc}{\ensuremath{{\mathrm{\,Me\kern -0.1em V\!/}c}}\xspace}
\newcommand{\gevcc}{\ensuremath{{\mathrm{\,Ge\kern -0.1em V\!/}c^2}}\xspace}
\newcommand{\mevcc}{\ensuremath{{\mathrm{\,Me\kern -0.1em V\!/}c^2}}\xspace}

\newcommand{\BABARPubYear}    {03}
\newcommand{\BABARPubNumber}  {030}
\newcommand{\SLACPubNumber} {10215}

\begin{document}

\preprint{BABAR-PUB-\BABARPubYear/\BABARPubNumber}
\preprint{SLAC-PUB-\SLACPubNumber}

\title{\boldmath Observation of a Narrow Meson Decaying to
$\Ds\piz\gamma$ at a Mass of 2.458~\gevcc}

%
\author{B.~Aubert}
\author{R.~Barate}
\author{D.~Boutigny}
\author{J.-M.~Gaillard}
\author{A.~Hicheur}
\author{Y.~Karyotakis}
\author{J.~P.~Lees}
\author{P.~Robbe}
\author{V.~Tisserand}
\author{A.~Zghiche}
\affiliation{Laboratoire de Physique des Particules, F-74941 Annecy-le-Vieux, France }
% ...
\collaboration{The \babar\ Collaboration}
\noaffiliation

\date{\today}

\begin{abstract}
A narrow state, which we label $\DsFE$, with mass 
$2458.0 \pm 1.0\;(\text{stat.}) \pm 1.0\;(\text{syst.})$~\mevcc,
is observed in the inclusive
$\Ds\piz\gamma$ mass distribution 
in 91~${\rm fb}^{-1}$ of
$e^+e^-$ annihilation data recorded by the \babar\  detector at
the \pep2
asymmetric-energy $e^+e^-$ storage ring. 
The observed width is consistent with the experimental
resolution. The data favor decay through $\DsTO\piz$ rather
than through $\DsTT\gamma$.
An analysis of $\Ds\piz$ data accounting for the influence of the
$\DsFE$ produces a $\DsTT$ mass of
$2317.3 \pm 0.4\;(\text{stat.}) \pm 0.8\;(\text{syst.})$~\mevcc.
\end{abstract}

\maketitle

Interest in the spectrum of charmed mesons has been heightened by the
discovery by this collaboration~\cite{Aubert:2003fg} of a narrow state,
produced in $e^+e^-\to c\overline{c}$ collisions at the
\pep2 collider,
decaying to $\Ds\piz$~\cite{cpfootnote},
with mass 2317~\mevcc, approximately 41 \mevcc below the
$DK$ mass threshold.
This state, the $\DsTT$, has been confirmed by CLEO~\cite{Besson:2003cp}
and Belle~\cite{Abe:2003jk,Abe:2003vu}.
Along with the $\DsTT$, we noted~\cite{Aubert:2003fg} the 
presence of
a narrow peak in the $\Ds\piz\gamma$ mass distribution near 
2.46~\gevcc. Because this signal is near the kinematic overlap of the
$\DsTT\gamma$ and $\DsTO\piz$ systems, special attention is
required to remove associated backgrounds and to distinguish between
the two possible decay modes.
Such an analysis is the subject of this paper. 

This state near 2.46~\gevcc has been seen by
CLEO~\cite{Besson:2003cp} and Belle~\cite{Abe:2003jk}
in the inclusive $\Ds\piz\gamma$ mass spectrum and 
by Belle~\cite{Abe:2003vu} in exclusively reconstructed 
$B$ decays.

To investigate the $\Ds\piz\gamma$ spectrum, we study
$\Ds$ candidates from $e^+e^-\to c \overline{c}$ 
(at a center-of-mass energy near 10.6~GeV)
that decay to $K^-K^+\pip$.
Particle identification
is used to provide clean samples of charged $K$ and $\pi$ candidates,
which are combined using a geometric fit to a common vertex.
Backgrounds are suppressed by selecting 
decays to $\Kbar^{*0}K^+$ and $\phi\pip$.
A description of this sample and
additional details can be found elsewhere~\cite{Aubert:2003fg}.  Events
with $1.954<m(\Km\Kp\pip)<1.981$~\gevcc are taken as $\Ds$ candidates.

A candidate $\piz$ is formed by constraining a pair 
of photons each with energy greater than 100~MeV
to emanate from
the intersection of the $\Ds$ trajectory
with the beam envelope, 
performing a one-constraint fit to the
$\piz$ mass, and requiring a fit probability 
greater than 5\%.  A given event
may yield several acceptable $\piz$ candidates. We retain only
those candidates for which neither 
photon belongs to another otherwise acceptable $\piz$.

Each $\Ds$ candidate is combined with all combinations of
accompanying $\piz$ candidates with momentum greater than 300~\mevc and
photon candidates of energy greater than 100~MeV. To suppress
background, photons that belong to any $\piz$ candidate
are excluded and we require the
momentum, $p^*$, of each 
$\Ds\piz\gamma$ combination
in the $e^+e^-$ center-of-mass frame to
be greater than 3.5~\gevc. The last
requirement also removes any $\Ds\piz\gamma$ combination from $B$ decay.

\begin{figure}
\includegraphics[width=\linewidth]{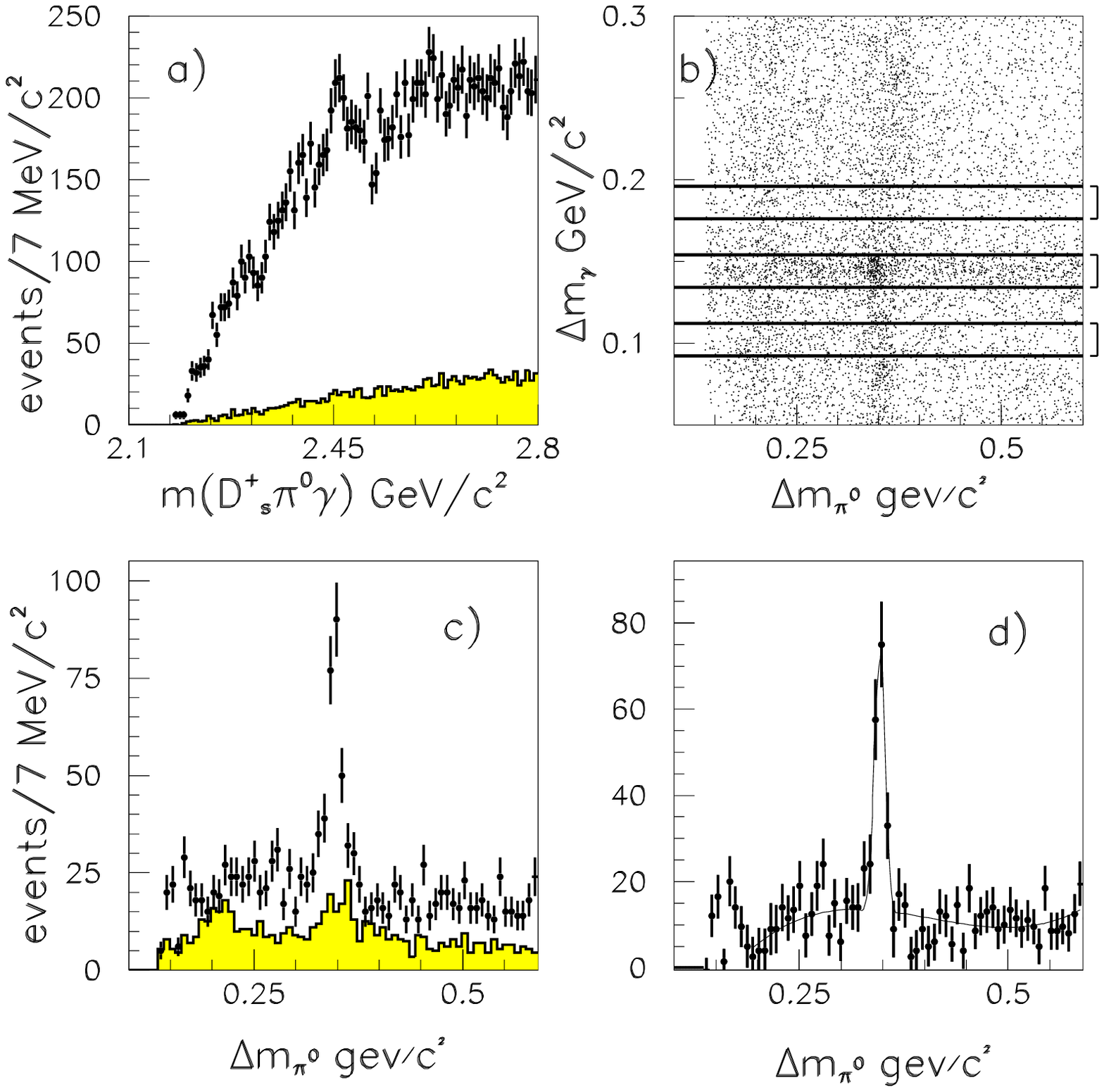}
\caption{\label{fig:dspig}
(a) The mass distribution for all selected $\Ds\piz\gamma$ 
combinations. The shaded region is from $\Ds$ sidebands defined by
$1.912<m(\Km\Kp\pip)<1.933~\gevcc$ and $1.999<m(\Km\Kp\pip)<2.020~\gevcc$.
(b) The value of $\DmDg$ versus $\DmDp$  for all combinations.
The horizontal lines delineate three
ranges in $\DmDg$. 
(c) The $\DmDp$ mass distribution for the middle
range of $\DmDg$ (points) and for the average of the upper and lower
ranges (shaded histogram).
(d) The difference between the two distributions shown in (c).
The curve is the fit described in the text. 
}
\end{figure}
The $\Ds\piz\gamma$ invariant mass distribution is shown in 
Fig.~\ref{fig:dspig}a. A clear enhancement is observed near 2.46~\gevcc.
The background underneath this peak is from several sources, which
can be described in terms of mass differences defined as
\begin{eqnarray}
\DmDg &\equiv& m(\Ds\gamma) - m(\Ds) \\
\DmDp &\equiv& m(\Ds\gamma\piz) - m(\Ds\gamma) \;.
\end{eqnarray}
A scatter plot of the data
is shown in Fig.~\ref{fig:dspig}b.
Particular background patterns are visible:
$\DsTO\to \Ds\gamma$ decay combined with an unassociated $\piz$,
which appears as a horizontal band,
and $\DsTT\to\Ds\piz$ decay combined with an unassociated $\gamma$,
which appears as a band that is almost vertical.

To demonstrate the existence of a signal above these backgrounds, 
the upper histogram of
Fig.~\ref{fig:dspig}c shows $\Ds\piz\gamma$ combinations
in the $\DsTO$ signal region, and the
gray histogram, scaled to the area of the signal region,
corresponds to the two $\DsTO$ sidebands.  We conclude that
a signal for a state decaying to $\Ds\piz\gamma$ exists over a 
background resulting from
$\DsTT$ and an unassociated $\gamma$. This background
peaks at a mass slightly higher than that of the signal.  A Gaussian fit
to the subtracted mass distribution (Fig.~\ref{fig:dspig}d)
indicates a narrow signal at $\DmDp=346.2\pm 0.9$~\mevcc
(statistical error only).

The state corresponding to this signal, 
which we label $\DsFE$, may decay to $\Ds\piz\gamma$ through
$\DsTO\piz$ or $\DsTT\gamma$.  To disentangle these
modes and reliably extract the parameters of the signal, we
apply an unbinned maximum likelihood fit simultaneously
to the $\Ds\piz\gamma$,
$\Ds\piz$, and $\Ds\gamma$ invariant masses of all $\Ds\piz\gamma$ 
combinations using the channel likelihood method~\cite{Condon:1974rh}. 
This fit describes the probability density function of
the two $\DsFE$ decay channels
as the product of a Gaussian shape in the $\Ds\piz\gamma$ 
mass distribution and a Gaussian shape projected into the 
$\Ds\piz$ or $\Ds\gamma$
mass axes, as appropriate. Because the daughter
resonances are narrow, 
interference between the two $\DsFE$ decay modes cannot be resolved,
and so is ignored.

Sources of background in the $\Ds\piz\gamma$ spectrum included in the fit are
purely combinatorial background ($\Ds$ meson combined with an unassociated 
$\piz$ and $\gamma$),
$\DsTO\to\Ds\gamma$ decay combined with an unassociated $\piz$, and
$\DsTT\to\Ds\piz$ decay combined with an unassociated $\gamma$. 
The fit also includes
a contribution from $\DsFE\to\DsTO\piz$ decay but with an unassociated $\gamma$
replacing the $\gamma$ from $\DsTO$ decay.
The fit determines the relative size of the background and signal
contributions, the mass and width of the $\DsFE$, and
the $\DsTT$ mass.

The likelihood fit is validated using Monte Carlo (MC) simulation.
This simulation includes $e^+e^-\to c\overline{c}$ 
events and all known charm states and decays, including the
$\DsTT$ and the signal under study.
The generated events were processed by a detailed 
detector simulation~\cite{Agostinelli:2002hh} and subjected to the same 
reconstruction and event-selection procedure as the data.

\begin{figure}
\includegraphics[width=\linewidth]{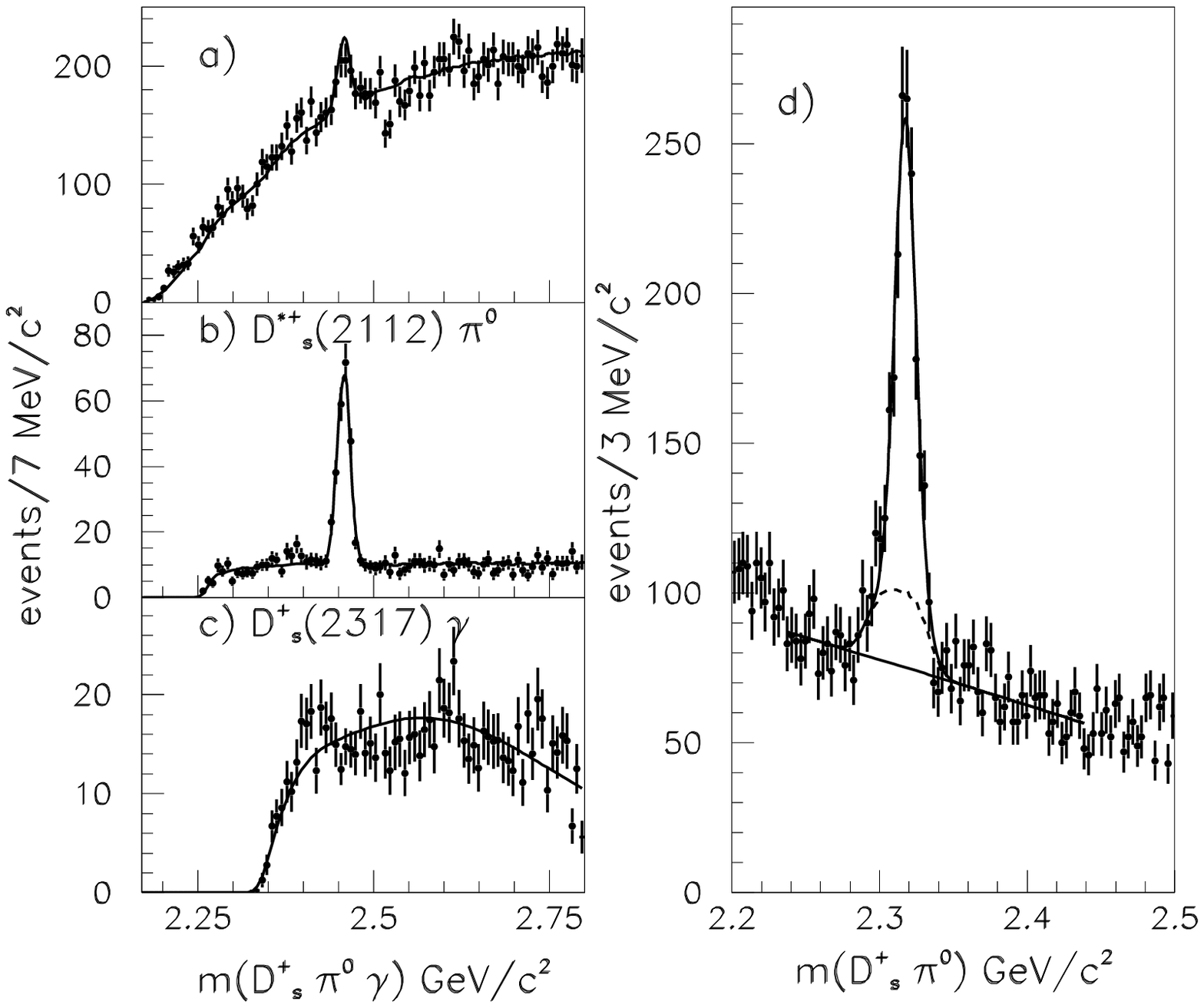}
\caption{\label{fig:fit}
Maximum likelihood fit results overlaid on the $\Ds\piz\gamma$ mass 
distribution with (a) no weights and 
after applying weights corresponding to
(b) the decay $\DsTO\piz$ and (c) the decay $\DsTT\gamma$.
(d) The mass spectrum of $\Ds\piz$ combinations (with no $\gamma$ requirement). 
The solid curve is the
fit described in the text. The dashed and lower solid curves
are the contributions from
$\DsFE$ decays and combinatorial background, respectively.
}
\end{figure}

As shown in Fig.~\ref{fig:fit}a, 
the fit provides a good description of
the $\Ds\piz\gamma$ mass distribution observed in the data.
The $\DsFE$ signal for a particular decay mode
can be isolated by calculating a weight for each $\Ds\piz\gamma$ combination
proportional to the relative likelihood contributed by the
decay mode of interest. Distributions of events so weighted
can be compared to the likelihood function to validate the fit.
This is shown in Figs.~\ref{fig:fit}b
and \ref{fig:fit}c. A $\chi^2$ probability calculation gives 
22\%, 74\% and 11\% for fig.~\ref{fig:fit}a, b and c respectively. 
The resulting yield of
correctly reconstructed $\DsFE\to\DsTO\piz$ ($\DsFE\to\DsTT\gamma$)
decays is  $195 \pm 26$ ($0 \pm 23$), consistent with the fit shown in 
Fig.~\ref{fig:dspig}d. Excluding the $\DsFE$ from the likelihood fit 
decreases the log
likelihood by approximately 57, corresponding to a significance of
more than $10$ standard deviations. 
The fit yields 
a $\DsFE$ mass of $2458.0 \pm 1.0$~\mevcc with an rms width
of $8.5 \pm 1.0$~\mevcc. 

\begin{figure}
\includegraphics[width=\linewidth]{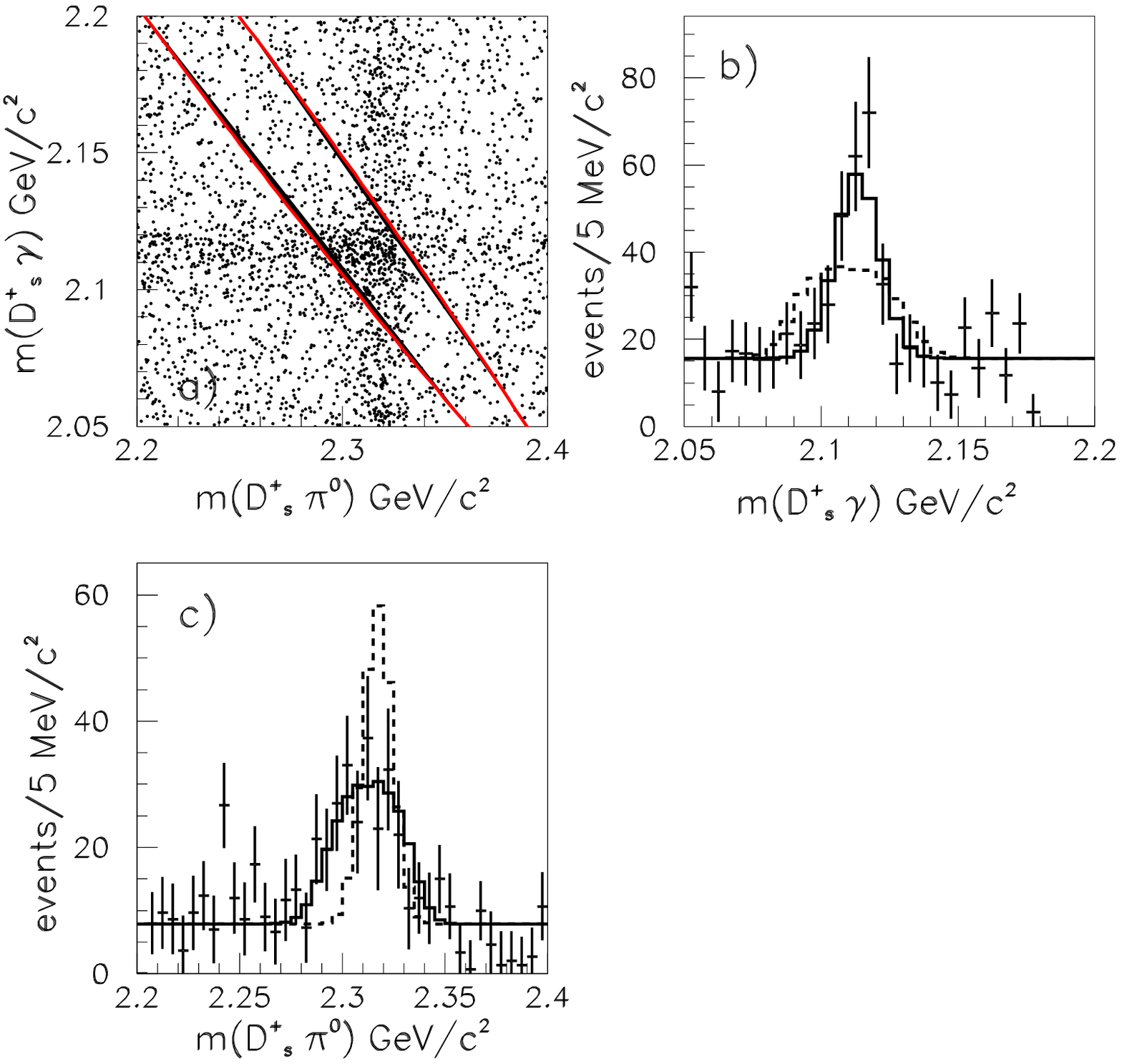}
\caption{\label{fig:proj}
(a) The $\Ds\gamma$ versus $\Ds\piz$ mass distribution for
all $\Ds\piz\gamma$ combinations. The decay of a zero-width $\DsFE$ 
is kinematically
restricted to the region between the two curves.
(b) Sideband-subtracted $\Ds\gamma$ mass distribution with
MC simulation for (solid histogram) $\DsFE\to\DsTO\piz$ and
(dashed histogram) $\DsFE\to\DsTT\gamma$. 
(c) A similar plot for the $\Ds\piz$ mass distribution.
}
\end{figure}

The likelihood fit uses the shapes of the $\Ds\piz$
and $\Ds\gamma$ mass distributions
to distinguish between the two possible decay modes, $\DsFE\to\DsTO\piz$ and 
$\DsFE\to\DsTT\gamma$.
These shapes are influenced by the kinematic constraints of
$\DsFE$ decay shown in Fig.~\ref{fig:proj}a.
Figs.~\ref{fig:proj}b--\ref{fig:proj}c show the
sideband-subtracted $\Ds\piz$ and $\Ds\gamma$ mass projections
compared with MC simulations of the two hypotheses (scaled to 
match the data yield).
The $\DsFE\to\DsTO\piz$ decay mode (solid histograms)
produces a narrow
$\Ds\gamma$ mass distribution and a wide 
$\Ds\piz$ mass distribution. 
In contrast, the $\DsFE\to\DsTT\gamma$
decay mode (dashed histograms) produces a wide
$\Ds\gamma$ mass distribution and a narrow 
$\Ds\piz$ mass distribution. Figures~\ref{fig:proj}b and \ref{fig:proj}c
show that the $\DsFE\to\DsTO\piz$ hypothesis is in better
agreement with the data.

Our previous measurement~\cite{Aubert:2003fg} of the $\DsTT$ mass
using the decay $\DsTT\to\Ds\piz$
did not explicitly consider background from $\DsFE\to\DsTO\piz$ decay.
This background peaks in the $\Ds\piz$
mass spectrum just below the $\DsTT$ mass. 
Shown in Fig.~\ref{fig:fit}d is the $\Ds\piz$ invariant mass distribution
for a sample of $\Ds$ candidates combined with all $\piz$ candidates,
with $p^* > 3.5$~\gevc. Superimposed on this distribution is
a binned fit that includes the contribution 
from the $\DsFE$ as estimated from MC simulation and
a quadratic background function.
The result is a $\DsTT$ yield
of $1022 \pm 50$ events, a mass of $2317.3 \pm 0.4$~\mevcc, and measured rms
width $7.3 \pm 0.2$~\mevcc. These results are 
an improvement over our earlier measurement~\cite{Aubert:2003fg}.

We divide the sources of systematic uncertainty in the $\DsFE$ and $\DsTT$
mass values and production rates into
three categories. The first category is associated with the fit procedure.
Likelihood fits to MC samples that include samples of
$\DsFE\to\DsTO\piz$ and $\DsFE\to\DsTT\gamma$ decays correctly reproduce
the given sample sizes within statistical errors. The average values of 
the fit results
obtained using statistically distinct MC samples corresponding
to the measurements in the data are used to place limits on any fit bias.

We obtain the background distribution
used in the likelihood from a random selection of $\Ds$, $\piz$, 
and $\gamma$ candidates taken from the MC
$\Ds\piz\gamma$ sample.
To test our sensitivity to this distribution, 
various selection requirements are altered 
within reasonable bounds
to provide alternate background samples for use in the fit. The resulting
changes in yield and mass are used as the second category
of systematic uncertainty.

Reconstruction of the decay sequences is the third source 
of systematic uncertainty.
To evaluate the reliability of
the MC determination of $\piz$ efficiency and momentum calibration,
we use control samples of $K_S \to \piz\piz$ and $\tau\to \piz X$.
On this basis, we assign
a systematic uncertainty of $\pm 5$\% in $\piz$ reconstruction
efficiency and a relative $\pm 1$\% in $\piz$ momentum bias.
Similar studies for $\gamma$ reconstruction reveal a systematic
uncertainty of $\pm 3$\% in $\gamma$ reconstruction efficiency
and $\pm 1$\% in energy bias.
Uncertainties in the $\Ds$ and $\DsTO$ masses,
taken from world averages~\cite{Hagiwara:2002pw},
also contribute to the systematic uncertainty.

The resulting total systematic uncertainty in the $\DsFE$ [$\DsTT$] mass 
is $\pm 1.0$ [$\pm 0.8$]~\mevcc.

Using the yields from our fit and correcting for efficiency,
we estimate the relative production rate
\begin{equation}
R=\frac{\sigma(\DsFE)\mathcal B(\DsFE\to\DsTO\piz)}{
\sigma(\DsTT)\mathcal B(\DsTT\to\Ds\piz)}
\label{eq:relyield}
\end{equation}
to be $0.25 \pm 0.03\;(\text{stat.}) \pm 0.03\;(\text{syst.})$,
requiring $p^*>3.5$~\gevc for both states.
We also estimate, at 95\% C.L.,
\begin{equation}
\frac{\mathcal B(\DsFE\to\DsTT\gamma)}{\mathcal B(\DsFE\to\DsTO\piz)} 
< 0.22\;.
\end{equation}

The observed rms width of the $\DsFE$ is consistent with detector resolution, 
as determined by Monte Carlo studies.
We conclude that the
intrinsic width of the $\DsFE$ is small ($\Gamma \lesssim 10$~\mevcc).

The mass of the $\DsFE$ lies above $DK$ and below $D^*K$ thresholds.
The narrow width and the isospin-violating decay to $\DsTO\piz$
indicate that decay to $DK$ is forbidden and
suggest an unnatural spin-parity assignment for the state.
Belle has observed the decay $\DsFE\to\Ds\gamma$ in production from both
$c\bar{c}$ continuum~\cite{Abe:2003jk} and $B$ decay~\cite{Abe:2003vu}.
Such a decay rules out $J=0$ and favors a $1^+$ interpretation.  
Decay
distributions studied by Belle further support $J=1$ for
$\DsFE$ and also $J^P=0^+$ for $\DsTT$.
The apparent absence of the decay $\DsFE\to\DsTT\gamma$ may indicate that the
electromagnetic decay mechanism cannot compete with $\DsTO\piz$, which 
may be a strong, but isospin-violating, process resulting from
$\eta$-$\piz$ mixing, as discussed by
Cho and Wise~\cite{Cho:1994zu}.

Our measurement of the $\DsFE$ mass
($2458.0\pm 1.4$~\mevcc, with combined statistical and systematic 
uncertainties) agrees with that obtained by Belle
($2456.5\pm 1.7$~\mevcc)~\cite{Abe:2003jk}, 
but is two standard deviations smaller than that obtained by CLEO
($2463.1\pm 2.1$~\mevcc)~\cite{Besson:2003cp}.  We obtain a relative yield 
($R=0.25 \pm 0.04$) which agrees with that of Belle 
($0.26 \pm 0.08$).
Both values are somewhat smaller than that reported by CLEO
($0.44\pm 0.13$).
Our reanalysis of the $\DsTT\to\Ds\piz$ sample to
account for background from the $\DsFE$ gives a
mass of $2317.3 \pm 0.4\;(\text{stat.}) \pm 0.8\;(\text{syst.})$~\mevcc,
which remains consistent with results from CLEO~\cite{Besson:2003cp}
and Belle~\cite{Abe:2003jk}.

In summary, in 91~${\rm fb}^{-1}$ of data collected
from the \babar\  experiment,
we have observed a narrow state that decays to $D_s^+\piz\gamma$
with a mass of 
$2458.0 \pm 1.0\;(\text{stat.}) \pm 1.0\;(\text{syst.})$~\mevcc.
The only significant $D_s^+\piz\gamma$ 
decay mode we observe is through $\DsTO\piz$.
We measure a mass and yield relative to the $\DsTT$ similar to those
measured by Belle though smaller than those reported by CLEO.
The observed width is compatible with our mass resolution. After including
the influence of this state, our new measurement of the $\DsTT$
mass is $2317.3 \pm 0.4\;(\text{stat.}) \pm 0.8\;(\text{syst.})$~\mevcc.

\begin{acknowledgments}
We are grateful for the 
extraordinary contributions of our \pep2\ colleagues in
achieving the excellent luminosity and machine conditions
that have made this work possible.
The success of this project also relies critically on the 
expertise and dedication of the computing organizations that 
support \babar.
The collaborating institutions wish to thank 
SLAC for its support and the kind hospitality extended to them. 
This work is supported by the
US Department of Energy
and National Science Foundation, the
Natural Sciences and Engineering Research Council (Canada),
Institute of High Energy Physics (China), the
Commissariat \`a l'Energie Atomique and
Institut National de Physique Nucl\'eaire et de Physique des Particules
(France), the
Bundesministerium f\"ur Bildung und Forschung and
Deutsche Forschungsgemeinschaft
(Germany), the
Istituto Nazionale di Fisica Nucleare (Italy),
the Foundation for Fundamental Research on Matter (The Netherlands),
the Research Council of Norway, the
Ministry of Science and Technology of the Russian Federation, and the
Particle Physics and Astronomy Research Council (United Kingdom). 
Individuals have received support from 
the A. P. Sloan Foundation, 
the Research Corporation,
and the Alexander von Humboldt Foundation.

\end{acknowledgments}


\begin{thebibliography}{9}
\bibitem[{cpf()}]{cpfootnote}
\bibinfo{note}{Inclusion of charge conjugate states is implied throughout this
  Letter.}
\end{thebibliography}

\end{document}